\documentclass[journal=jpcafh,manuscript=article]{achemso}

\usepackage{chemformula} 
\usepackage[T1]{fontenc} 
\usepackage{braket}
\usepackage{amsmath}
\usepackage{color}
\usepackage{ulem}

\author{Henning Kirchberg}
\affiliation{I.Institut f\"ur Theoretische Physik, Universit\"at Hamburg, Jungiusstr. 9, 20355 Hamburg, Germany}
\author{Peter Nalbach}
\affiliation[]{Westf\"alische Hochschule, M\"unsterstr. 265, 46397 Bocholt, Germany}
\author{Christian Bressler}
\affiliation[]{Center for Free-Electron Laser Science, DESY and The Hamburg Centre for Ultrafast Imaging, Universit\"at Hamburg, 22607 Hamburg, Germany}
\alsoaffiliation[]{European XFEL, Holzkoppel 4, 22869 Schenefeld, Germany}
\author{Michael Thorwart}
\email{michael.thorwart@physik.uni-hamburg.de}
\affiliation[]{I.Institut f\"ur Theoretische Physik, Universit\"at Hamburg, Jungiusstr. 9, 20355 Hamburg, Germany}

\title{Spectroscopic Signatures of the Dynamical Hydrophobic Solvation Shell Formation}

\abbreviations{IR,NMR,UV}

\begin{document}


\begin{abstract}
When a hydrophilic solute in water is suddenly turned into a hydrophobic species, for instance, by photoionization, a layer of hydrated water molecules forms around the solute on a time scale of a few picoseconds. We study the dynamic build-up of the hydration shell around a hydrophobic solute on the basis of a time-dependent dielectric continuum model. Information about the solvent is spectroscopically extracted from the relaxation dynamics of a test dipole inside a static Onsager sphere in the nonequilibrium solvent.  The growth process is described phenomenologically within two approaches. First, we consider a time-dependent thickness of the hydration layer which grows from zero to a finite value over a finite time. Second, we assume a time-dependent complex permittivity within a finite layer region around the Onsager sphere. The layer is modeled as a continuous dielectric with a much slower fluctuation dynamics.  We find a time-dependent frequency shift down to the blue of the resonant absorption of the dipole, together with a dynamically decreasing line width, as compared to bulk water. The blue shift reflects the work performed against the hydrogen bonded network of the bulk solvent and is a directly measurable quantity. Our results are in agreement with an experiment on the hydrophobic solvation of iodine in water. 
\end{abstract}

\section{Introduction}
Solvation dynamics is one of the fundamental processes in nature and plays a central role in many chemical reactions and charge transfer processes \cite{NitzanBook,ReviewElsasser17,PerspectElsasser17, ReviewBagchi10,ReviewBagchi00, MennucciBook08,BagchiBook12}. An important focus at present is to understand the time-dependent interplay of a polar solvent and a changing charge distribution in a solute molecule whose time scales range in the few-10 fs to the few-ps regime.\cite{mon2017} Water is certainly the most prominent solvent in nature and shows particular features due to the structure of the water molecule and the resulting ability to form strong hydrogen bonds among each other, but also with the solute, if it has a net charge or polar groups. By introducing an apolar solute, the hydrogen bonds are more or less broken while no hydrogen bonds to the solute are formed. Both scenarios lead to a structural rearrangement of the water molecules in close proximity to the solute as compared to the bulk.

A useful concept is the notion of the hydration shell formed around the solute \cite{ReviewElsasser17}. Qualitatively, it consists of the first layer of water molecules which directly surround the solute and usually, not more than a few (up to five  \cite{ReviewElsasser17}) following layers exist (if at all). The features that distinguish hydration layers may depend strongly on the property measured. While single particle properties like diffusion (both translation and rotation) are modified depending on the location of the water molecule, collective properties (like dielectric relaxation and solvation dynamics) could be more extensive. Dielectric relaxation (if can be decomposed layer-wise) is found to extend to the third or even fourth layer. \cite{ReviewElsasser17, PerspectElsasser17} Neutron scattering experiments find signatures around a protein even beyond $20$ {\AA} (beyond 5 layers). The fastest fluctuations produced by the water in the first few layers occur on a time scale of a few hundred fs and are mainly due to the librational motion of the water molecules which are hindered in their free rotation. The slowing down of these fluctuations as compared to the bulk is a matter of discussion, but seems not to exceed a factor of 5 to 10 compared to the time scale of the bulk fluctuations (we do not consider here peculiar situations when water molecules are trapped in grooves of DNA or in clefts of protein surfaces) \cite{ReviewElsasser17}.  Overall, the structured water in the hydration layers become somewhat more rigid, but the extreme picture of the solvated molecule as an ice-berg swimming in the bulk water turns out to be inappropriate \cite{ReviewElsasser17,ReviewBagchi10,BagchiBook12}.

To describe the solvation dynamics theoretically, different types of approaches exist. On the one hand, more microscopic approaches have been taken in molecular hydrodynamic calculations \cite{Halle03,ReviewBagchi10} and in molecular dynamics simulations \cite{Laage14,ReviewElsasser17}. Furthermore, solvation processes may be directly calculated using the chemical potential of introducing the solute into the solvent based on the two- or three-particle radial distribution functions \cite{gar1996,sed2011}. On the other hand, there are more phenomenological approaches build on models with a dielectric continuum \cite{ReviewBagchi10,ReviewBagchi00, MennucciBook08,BagchiBook12,NeeZwanzig1970}.All directions have their advantages and limitations. The molecular picture is more accurate and commonly closer to the experiment, but may become numerically expensive, especially when quantum mechanical effects like hydrogen bonds have to be included. The approach on the basis of continuum models is conceptually simple and intuitive and even allows for analytic solutions. It may give a better qualitative understanding, but naturally is often more inaccurate. Yet, Bagchi, Oxtoby and Fleming developed a continuum theory which explains with high accuracy the time-dependent Stokes shift of a molecule in a polar medium.\cite{bag1984}

Two examples of important solvation effects in water are hydrophobic and hydrophilic solvation \cite{BagchiBook12,BagchiBook13}.  
A hydrophilic molecule likes water molecules and interacts with them strongly and directly, mainly electrostatically. 
The hydrophilic objects are polar or charged and water molecules can easily form hydrogen bonds with them, such that 
the network of the hydrogen bonded water is substantially distorted near the solute. Hydrophobic hydration describes the interaction between one non-polar solute and the surrounding water molecules. Hydrophobic solvation is considered to be the major driving force behind diverse phenomena from the creation of micro-emulsions to make new materials to fundamental biological protein folding processes \cite{pra2002,dil2005}.  The hydrophobic effect is multifaceted and is a complex phenomenon, and is not fully understood yet. Albeit its great importance in numerous processes, the origin of hydrophobicity, especially from the microscopic viewpoint on atomic scale, remains a highly disputed topic in physics, chemistry and biology \cite{grd2017,cha2005,blo1993,ReviewElsasser17}. 

The process of hydrophobic hydration can be illustrated on the basis of thermodynamic arguments  \cite{blo1993} and is related to a  positive change of the Gibbs free energy. This increase of energy is directly associated with the poor solubility of the hydrophobic solute. To quantify this energetic penalty, one considers the two contributions to the change of Gibbs free 
energy, $\Delta G=\Delta H - T\Delta S$, where $\Delta H$ is the change of enthalpy and $\Delta S$ the entropic change.
The solvation process involves a disruption of the hydrogen bond network of the water molecules and the refilling of the resulting vacancy with the apolar component. Due to the insertion of the hydrophobic solute,  small solute-solvent interactions are formed out, particularly by weakened Van-der-Waals forces, leading to an enthalpic gain, i.e., a negative $\Delta H$.
The water molecules in the vicinity of the apolar solute rearrange themselves, create enhanced hydrogen bonds with each other and form out an enhanced network to minimize the disruption of the hydrogen bonds by the apolar molecule. By this, the rearranged water molecules are restricted in their mobility as compared to before \cite{has1995} and increase their spatial order compared to the highly fluctuating bulk water\cite{fra1945}. 
This reduces entropy, i.e., implies  a negative $\Delta S$ which is larger than the enthalpic gain $\Delta H$, such that the total free energy for the apolar molecule with the additional enhanced hydration layer is positive. The free energy remains positive for a large range of temperature. Its temperature dependence is such that it is entropic (negative $\Delta S$) in a narrow temperature range around room temperature, 
 but becomes enthalpic (positive $\Delta H$) at higher temperatures. Hence, the overall positive free energy signals a poor solubility for the apolar solute.

This thermodynamic picture of hydrophobic solvation has been confirmed by atomistic molecular dynamics simulations of the hydration of ions in aqueous solution at room temperature \cite{lyn1997,pas2004}. 
Neria and Nitzan \cite{Nitzan92} have carried out a detailed computer simulation study 
of solvation dynamics in ionic solutions.  A fast Gaussian decay of the solvent response has been found which is 
followed by a very slow exponential-like decay. It was also found that the Gaussian time constant is
practically independent of the ion concentration whereas the long
time exponential time constant depends rather strongly on it, and, in fact,  decreases when the ion concentration is
increased.

For the hydrophobic solvation there is a large range of time scales, ranging from ultrafast collective polarization of the bulk water on the sub-100 $\rm fs$ time scale up to orientational reorganization within the hydration shell occuring within several $\rm ps$. Mondal, Mukherjee and Bagchi provide a detailed overview about all temporal contributions with a molecular dynamics simulation \cite{mon2017}. Different experimental techniques are more or less sensitive to the respective time scales of the dynamics. NMR studies provide, for example, the single particle dynamics but averaged over a large number of molecules\cite{mat2008}, while dielectric relaxation spectroscopy sums over a collective contribution from all the molecules \cite{nan1997,sed2011}.

X-ray absorption spectroscopy (XAS) is able to probe the transition form hydrophilic to hydrophobic solvation and the dynamic built-up of the solvation shell with atomic scale resolution on a short time scale \cite{pha2011}.
The process of the formation of a solvation shell is usually observed if the central molecule abruptly changes its charge configuration, e.g., from its ionic to its neutral, and hence, hydrophobic state. This can be achieved by a strong-field removal of one electron of the solute. The  picosecond XAS study reveals the creation of a network of hydrogen bonded water molecules in the first solvation shell within 5$\rm ps$ around a iodine molecule which has been neutralized just before. The experimental data further demonstrate the dynamic expansion of the solvent layer after the transfer $I^- \to I^0$ with an increase of the radius of the 
hydrated cluster by up to $80\%$ \cite{pha2011,kon1998}. The experimental data have been corroborated by ab initio molecular dynamics simulations. Further experimental studies have been carried out, see Ref.\ \cite{ReviewElsasser17} for an excellent recent account on the present status. 

In this work, we study the time-dependent formation of the hydration shell around a hydrophobic solute on the basis of a time-dependent dielectric continuum model. We determine the time-dependent response of a test dipole in a vacuum Onsager spherical cavity. The sphere is placed in a dielectric continuum around which a dielectric layer of hydrated water grows on a finite time scale.  The growth process is imposed from outside and described phenomenologically within two approaches. First, we consider a time-dependent thickness of the hydration layer which grows from zero to a finite value over a finite time range. Second, we assume a time-dependent variation of the dielectric properties in the form of a time-dependent complex permittivity within a finite layer around the central Onsager sphere. In the first approach, we assume that the hydration shell has different dielectric properties and, in particular, a slower reorganization (or, Debye) time.  We find a time-dependent frequency shift down to the blue of the resonant response of the dipole, together with a dynamically decreasing line width. The blue shift directly indicates the work which the systems performs to form the hydration layer and is a directly measurable quantity. The reduced line width reveals the less effective damping of the hydration shell which has slower fluctuations and also removes the fast bulk fluctuations further away from the central dipole.

\section{Dynamic formation of the hydration shell}
To describe the impact of a time-dependent formation of the hydration shell on the relaxation properties of a test dipole dissolved in a polar solvent, we generalize the static Onsager model  \cite{ons1936,kir2018,NeeZwanzig1970} to include also time-dependent nonequilibrium effects. Its dipole moment is assumed to be weak and placed in the center of a vacuum  Onsager spherical cavity with fixed radius $a$. The dipole moment reacts to an externally applied electric field to study its frequency-dependent response  during the dynamic formation of the hydration shell formation, being a highly non-equilibrium situation. To mimick the growth of the hydration shell, we assume a dynamic layer thickness grwoing from zero to a final value $r$. The precise origin of this growth is left unspecified and only enters phenomenologically into the model. Outside the shell, the water shows its usual bulk behavior. This model is motivated by the experiment \cite{pha2011} with a negatively charged hydrophilic iodide which is ionized by an X-ray pulse. Around the resulting neutral and hydrophobic iodine, an expanded water layer forms containing around 27 water molecules \cite{kon1998}. The extracted size of the entire complex, i.e., the central iodine plus the hydration shell of one layer of water molecules after the shell formation is approximately $b=6.5${\AA}. We extract this value from  the iodine-oxygen radial distribution function, where a second maximum appears signaling a high density of water atoms beyond the shell and the countinuous interface to the bulk water\cite{kon1998}. The initial size of the dissolved iodide is estimated as $a=3.6${\AA} from the first maximum of the iodide-oxygen radial distribution. The resulting layer thickness after neutralization can thus be obtained as $r=2.9${\AA}$=0.8 a$. We neglect in our model the hydration shell around the initial iodide formed by approximately  eight water molecules in the first solvation shell.\cite{mck2005,kon1998} We study two different scenarios for the hydration shell formation. Within the first one (A.), we consider a time-dependent growth of a hydration layer with a time-dependent thickness, which has a  complex permittivity different from the bulk. In the second situation (B.), we assume a static Onsager sphere, but a  time-dependent complex permittivity within a well-defined spatial range associated with the fixed shell extension. Outside the hydration layer, bulk water is assumed as solvent. 
In comparing the results of both scenarios we may conclude interesting insights into the dynamic formation of a hydration  shell leading to the hydrophobic behavior. 
 
\subsection{A. Increasing radial shell extension}
In the Onsager model, the solvent is described as continuous, homogeneous and uniform and is associated with the time translationally invariant dielectric function $\epsilon_x(t-t')$ \cite{mck2005}. The index $x$ refers to different components of the environment, i.e., the hydration shell ($x=H$) and the bulk ($x=B$).  Throughout this work, we consider the Debye form of the dielectric function
\begin{align}
\epsilon_x(t-t')=\epsilon_{x,\infty}\delta(t-t')+\frac{\epsilon_{x,s}-\epsilon_{x,\infty}}{\tau_{x,D}}\exp\bigg[-\frac{t-t'}{\tau_{x,D}}\bigg]\Theta(t-t'),
\label{EQ1}
\end{align}
where each component $x$ has its own specific Debye relaxation time $\tau_{x,D}$, its static dielectric constant $\epsilon_{x,s}$ and its high-frequency dielectric constant $\epsilon_{x,\infty}$. For bulk water at room temperature, $\epsilon_{B,s}=78.3$, $\epsilon_{B,\infty}=4.21$, and $\tau_{B,D}=8.2 \rm ps$\cite{mck2005,buc1999}. The dielectric properties of the hydration shell continuum are less clear, but in general, a higher structural order of the water molecules in the first few layers around the solute implies weaker fluctuations and, thus,    
$\epsilon_{B,s}\gg \epsilon_{H,s}$. \cite{mck2005,ReviewElsasser17,PerspectElsasser17,tie2010}. This reflects the fact, that the water molecules are stronger bound in a hydrogen network and are less polarizable. Due to this enhanced interaction,  the relaxation time is slower by a factor of 5 to 10 \cite{ReviewElsasser17}, such that $\tau_{H,D}\gg \tau_{B,D}$ \cite{tie2010}. A reduced dielectric constant of water is also recorded in water strongly confined in small tubes of several nanometers \cite{fum2018}. Here, water exhibits a less flexible structure near surfaces and is difficult to reorient by applying an electric field. 

We assume that the shell formation process is phenomenologically well-described by a radial increase of the total radius  $b(t)$ of the central Onsager sphere plus layer of the form
\begin{align}
\label{EQ2}
b(t)=a+r[1-e^{-\alpha (t-t_0)}]\Theta(t-t_0),
\end{align}
where the shell formation begins at time $t_0$  from the static radius $a$ with the rate $\alpha$ of formation. The final shell thickness is $r$ and the final total radius is $b=a+r$. 
This form of the growth reflects a gradual shell formation, where the environmental dipole moments adjust themselves first more rapidly layer by layer in a stronger bound network and then successively slower if further away from the central hydrophobic molecule. 

\subsection{Reaction field}
To determine the time-dependent response of a dipole to the dynamic hydration shell formation, we need to calculate the resulting electric field induced by the polar environment.
The central time-dependent dipole moment polarizes both components of the polar environment, the hydration shell and the bulk water, such that both create a fluctuating reaction field.  
We assume that the dipole moment does not change its direction but its magnitude. 

Due to the spheric symmetry, the electric potentials in the cavity, the shell and the bulk water are expressed in terms of Legendre polynomials \cite{bot1973}. From the potentials,
 all resulting electric fields in all media result by imposing the boundary conditions at $a$ and $b$, i.e.,  (i) the continuity of the electric potentials and (ii) the continuity of the normal components of the electric displacement. The boundary conditions follow as
\begin{align}
\label{EQ3}
R(t)a^3+\mu(t)&=A(t)a^3+B(t) \\ 
\label{EQ4}
R(t)a^3-2\mu(t)&= \int_{t^*}^t dt' \epsilon_H(t-t')\bigg[A(t')a^3-2B(t')\bigg] \\
\label{EQ5}
A(t)b(t)^3+B(t)&=C(t) \\
\label{EQ6}
\int_{t^*}^t dt' \epsilon_H(t-t') \bigg[A(t')b(t)^3
-2B(t')\bigg]&=-2\int_{t^*}^t dt' \epsilon_B(t-t')C(t')\, ,
\end{align}
where $R(t)$ is the magnitude of the reaction field and $\mu(t)$ the magnitude of the dipole moment. $A(t)$ is the reaction field and $B(t)$ the field of the induced dipole moment in the shell, while $C(t)$ is the induced dipole field in the bulk water. We have to ensure that the formation is slow enough to disregard any dynamic magnetic field induced by the displacement current. We further disregard the spatial variation of the dielectric function, but assume it  homogeneous in the bulk and the shell. The time $t^*$ marks the beginning of measuring the response of the dipole moment, when the molecular charge is brought out of its equilibrium, and may differ from the time $t_0$ of the beginning of the shell formation.
The two processes may be recorded experimentally by two different sharp coherent pulses, like in an experimental pump-probe set-up.

When performing a Fourier transform $\mathcal{F}[f(t)]=\int_{t^*}^\infty dt \exp(i\omega t) f(t)$ of Eqs.\ (\ref{EQ3}) and (\ref{EQ4}),  we want to treat the time-dependent total radius $b(t)$ adiabatically. This is possible as long as the shell formation  in  Eq.\ (\ref{EQ2}) is much slower than all time-decaying processes of the electric fields induced by the environment, or, equivalently, 
the rate $\alpha$ is the smallest of all inverse time scales of the problem. In doing so, $b(t,t_0)^3\to b(t^*,t_0)^3$ contributes as a constant parameter and can be taken out of the Fourier integral which starts effectively at time $t^*$, while the shell formation has already started at $t_0$. We require $t^*\geq t_0$ such that the beginning of the shell formation is at the same time or before the beginning of the measurement. Below, we will ensure self-consistently that $\alpha$ fulfills this requirement. 
From Eqs.\ (\ref{EQ3})-(\ref{EQ6}), we obtain for the reaction field in Fourier space
\begin{align}
\label{EQ7}
R(\omega)&=\frac{\mu(\omega)}{a^3}\frac{-2[\epsilon_H(\omega)+2][\epsilon_H(\omega)-\epsilon_B(\omega)]a^3+2[\epsilon_H(\omega)-1][2\epsilon_B(\omega)+\epsilon_H(\omega)]b(t_0,t^*)^3}{2[\epsilon_H(\omega)-1][\epsilon_H(\omega)-\epsilon_B(\omega)]a^3-[2\epsilon_B(\omega)+\epsilon_H(\omega)][2\epsilon_H(\omega)+1]b(t_0,t^*)^3} \\ \notag
&=\xi(\omega,t_0,t^*)\mu(\omega),
\end{align}
where $b(t^*,t_0)^3$ enters parametrically and the dipole moment may be understood as $\mu(\omega)=eq(\omega)$. This relation defines the frequency-dependent 
susceptibility $\xi(\omega,t_0,t^*)$ which depends parametrically on $t_0$ and $t^*$.

\subsection{Equation of motion for the molecular dipole moment}
We study next the time-dependent expectation value $\braket{\mu(t)} =e \braket{ q(t)}$ of the molecular dipole moment. The net charge of the molecule is zero, but it can be polarized such that within the harmonic approximation, the charge oscillates with a characteristic frequency $\omega_0$ around its equilibrium position, forming the time-dependent molecular dipole moment. For  $\braket{q(t)}$, we can then derive an equation of motion while we take into account the  reaction field, the environmental back action of the bulk water and the gradually growing hydration shell given in Eq.\  (\ref{EQ7}), which couples to the dipole moment with the force $e\braket{R(t)}$. Furthermore the polarized charge couples to an additional external electric probe field associated with the force $eE(t)$.
The resulting integro-differential equation of motion follows as 
\begin{align}
\label{EQ8}
m&\braket{\ddot{q}(t)}+m\omega_0^2 \braket{q(t)}-e^2 \int_{t^*}^t dt' \xi(t^*,t_0,t') \braket{q(t')}=eE(t),
\end{align}
with the electron mass $m$, the electron charge $e$ and a typical molecular frequency $\omega_0=2.5\times10^{15}\rm{Hz}$ in the near ultraviolet spectrum. 
Moreover, all information about the molecular extension $a$, the final hydration shell thickness $r$, its formation rate $\alpha$ and the time $t_0$ of the formation beginning enter via the Fourier transform of $\xi(\omega,t_0,t^*)$, obtained from Eq.\ (\ref{EQ7}). 
In the equation of motion (\ref{EQ8}), $\operatorname{Re}[\xi(t^*,t_0,t')]$ induces a renormalization of the potential frequency, while $\operatorname{Im}[\xi(t^*,t_0,t')]$ leads to  damping. A Fourier transform of Eq.\ (\ref{EQ8}) leads to
\begin{align}
\label{EQ9}
\braket{q(\omega)}=\frac{ \braket{\dot{q}(t^*)}e^{i\omega t^*}-i\omega \braket{q(t^*)}e^{i\omega t^*}+(e/m) E(\omega)}{\omega_0^2-\omega^2-(e^2/m) \xi(\omega,t_0,t^*)},
\end{align}
with the initial charge displacement $\braket{q(t^*)}$ and its initial velocity $\braket{\dot{q}(t^*)}$, while $\braket{q(t\to \infty)}=\braket{\dot{q}(t\to \infty}=0$ as the functions are integrable.

To obtain the electric reaction field $A(t)$ and the dipole field $B(t)$ in the hydration shell, we transform $\braket{q(\omega)}$ in Eq.\ (\ref{EQ9}) back into the time domain. By this, we obtain the dipole moment $\braket{\mu(t)}=e\braket{q(t)}$ and the resulting reaction field given in Eq.\ (\ref{EQ7}) in the Onsager sphere and can then calculate the associated fields $A(t)$ and $B(t)$ as given in Eqs.\ (\ref{EQ3}-\ref{EQ6}). This allows us to verify self-consistently that we can treat $b(t,t_0)$ in Eqs.\ (\ref{EQ5}) and (\ref{EQ6}) adiabatically. 
We are free to choose the initial time $t^*$, when the dipole begins to oscillate, while the hydration shell formation has begun at an earlier time $t_0$, i.e.,  $t^*\geq t_0$. We set $t_0=0$ such that $t^*$ directly refers to the time span during which the hydration shell grows.  Additionally, we switch off the driving field $E(\omega)=0$ to see the damping induced by the environment, set  $\braket{\dot{q}(t^*)}=0$ and choose the typical oscillator length for the initial expectation value for the charge displacement as $\operatorname{Re}[\braket{q(t^*)}e^{i\omega t^*}]=\sqrt{\frac{\hbar}{m \omega_0 }}$. 
\begin{figure}[h!!!]
\centering
\includegraphics[scale=0.55, angle=-90]{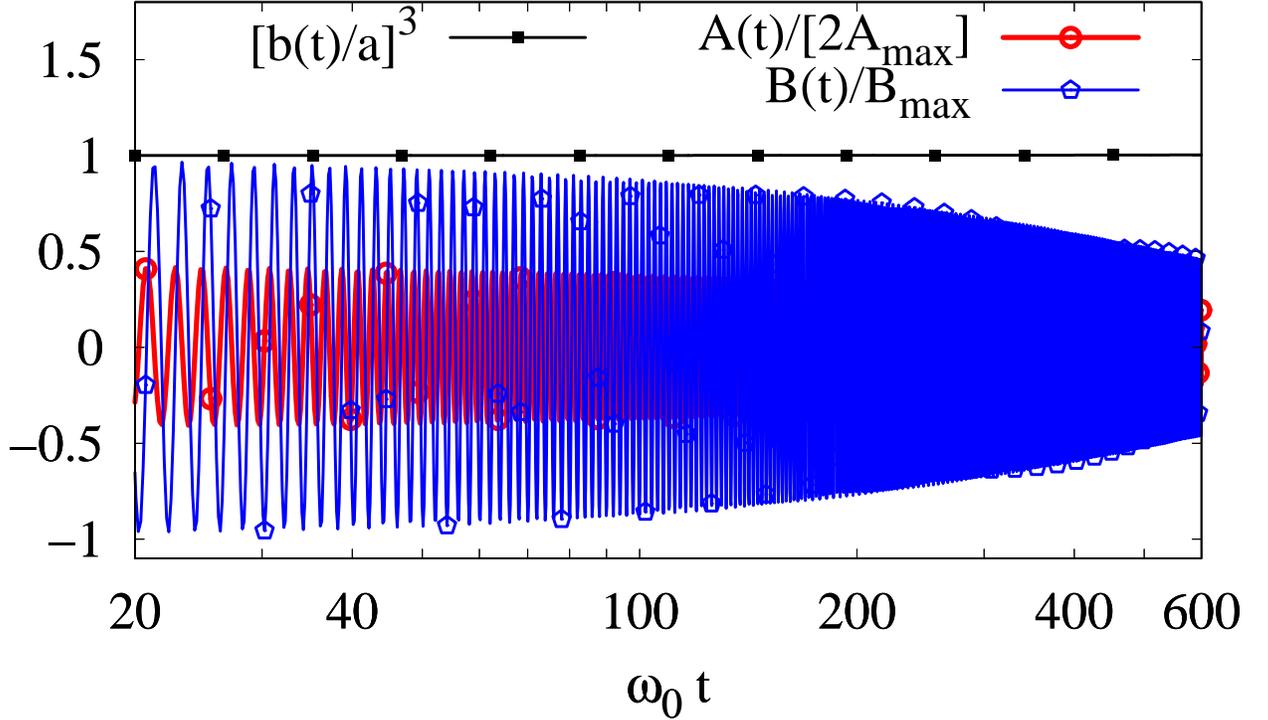}
\caption{Dipole field $A(t)$ and reaction field $B(t)$ in the hydration shell which has begun to grow at $t_0=0$ before we start the observation at $t^*=10\omega_0^{-1}$. The inner (constant) sphere radius is $a=3.6$ {\AA}, while the final shell thickness is set to  $r=0.8a$. $\Gamma=2.6 \times 10^{-4}\omega_0$ is the resulting decay rate of the electric fields. The formation rate is chosen as $\alpha=0.1\Gamma$ and the Debye dielectric parameters are set to $\epsilon_{H,\infty}=\epsilon_{B,\infty}$, $\epsilon_{H,s}=0.1 \epsilon_{B,s}$ and $\tau_{D,H}=10 \tau_{D,B}$. \label{FIG1}}
\end{figure}
The results are shown in Fig.\ \ref{FIG1}. Form this, it is clear  that  $b(t)^3$ increases slowly enough in comparison to $A(t)$ and $B(t)$ for a proper choice of the formation rate $\alpha$ such that the time-dependence of the total radius can be treated adiabatically in Eqs.\ (\ref{EQ5}) and (\ref{EQ6}), i.e., $b(t)^3\equiv b(t^*)^3=$const. 
We also have to ensure that the resulting damping has to comply with $\Gamma\gg \alpha$ for different sphere radii $a$, all given final hydration shell thicknesses $r$ and for all chosen times $t^*$ for the initial displacement. For the specific choice of the neutralization of iodine ($I^-\to I^0$) with $a=3.6${\AA} and $r=0.8a$, we  find  $\Gamma=2.6\times 10^{-4}\omega_0$, where the oscillation begins at $t^*=10\omega_0^{-1}$ after the hydration shell formation. We choose $\alpha=0.1\Gamma$, which refers to a typical time scale of the hydration shell formation of $\alpha^{-1} \sim \rm ps$\cite{pha2011,PerspectElsasser17}. The shell formation requires a hydrogen-bond breaking of the solvent molecules with the just neutralized and hydrophobic iodine and a reformation of strong hydrogen-bonds  among each other. The measured average molecular reorientation times for bulk water is $2.5 \rm ps$. \cite{PerspectElsasser17,bak2008}

\begin{figure}[h!!!]
\centering
\includegraphics[scale=0.55, angle=-90]{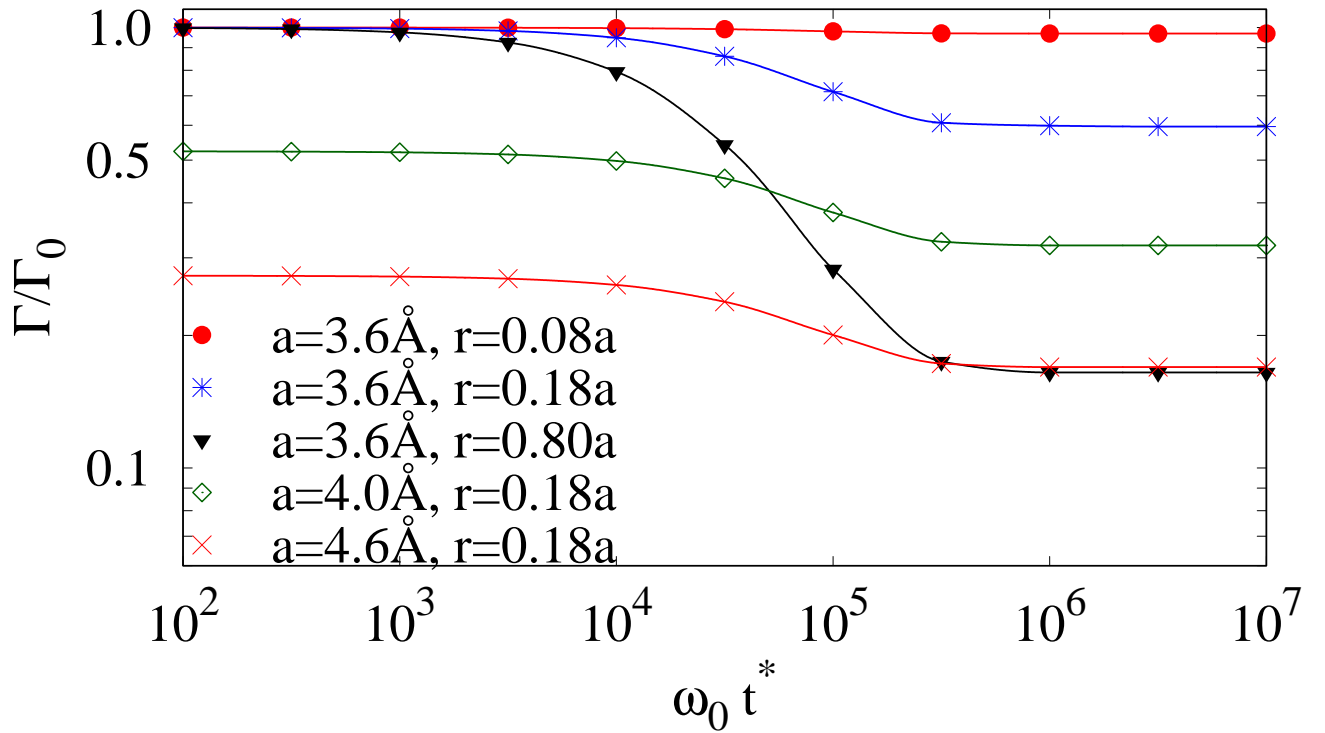}
\caption{Damping rate in dependence of the time $t^*$ between the start of the shell formation and initial displacement for different sphere radius $a$ and final shell extensions $r$. The damping is recorded relative to the damping $\Gamma_0=2.6\times 10^{-4} \omega_0$ for $a=3.6${\AA} and $r=0.8a$ at $t_0=0$. Moreover, $\epsilon_{H,\infty}=\epsilon_{B,\infty}$, $\epsilon_{H,s}=0.1 \epsilon_{B,s}$ and $\tau_{H,D}=10 \tau_{B,D}$.\label{FIG2}}
\end{figure}

The dynamics of the displacement of the central dipole is damped with the damping rate $\Gamma$. The results for $\Gamma$ 
 for different given parameters are shown for increasing time $t^*$ in Fig.\ \ref{FIG2}. An overall decrease of the damping is observed with increasing time delay $t^*$ between the beginning of the shell formation and the initial charge displacement. 
The fact that $\Gamma$ is time-dependent reflects the nonequilibrium situation of the time-dependent environment with the growing hydration layer.  The maximal damping rate is observed at $t^*=0$, where no shell exists. While the shell forms out, the bulk water as the source of fluctuations is further pushed away from the central dipole by the intermediate hydration layer. The dielectric properties within the newly created hydration shell are assumed to be weakened as the water molecules hold stronger together due to the formation of hydrogen bonds. This reduces damping of the charge displacement and hence of the induced electric fields $A(t)$ and $B(t)$.  As can be seen, for larger inner sphere radii $a$, the damping decreases as the environment is spatially further away from the central dipole, see Fig.\ \ref{FIG2}. Qualitatively similar results have been obtained previously for the relaxation time of an excited vibrational state of a molecular complex with a finite, but static hydration shell around \cite{nal2014}. If the hydration shell grows to a bigger final thickness $r$, damping is more reduced as the strongly fluctuating bulk water is gradually replaced by the less polarizable water confined in the hydration shell. The inverse effect is recorded for a smaller final shell extension $r$ in Fig.\ \ref{FIG2}. 

\subsection{Response function}
As we are only interested in the response of the dipole moment to the external electric field, we set the initial conditions $\braket{q(t^*)}=0$ and $\braket{\dot{q}(t^*)}=0$ in Eq.\  (\ref{EQ9}). The external electric field drives the charge, which follows with a characteristic but fixed phase delay and amplitude difference after a transient time. On the one hand, the pulse has to be long enough that the dipole displacement follows its stationary oscillation, but on the other hand, it should be short enough that the time-dependent total radius $b(t)\equiv b(t^*)$ can be treated adiabatically. This can expressed as $\Gamma_{\rm min}^{-1}<T<\alpha^{-1}$, where $T$ is the pulse duration. $\Gamma_{\rm min}$ is the minimal damping which occurs when the shell formation process is completed (Fig.\ \ref{FIG2}). The resulting response to the external driving field can be evaluated as 
\begin{align}
\label{EQ10}
\braket{q(\omega)}=\frac{e}{m}\frac{1}{\omega_0^2-\omega^2-(e^2/m) \xi(t^*,\omega)}E(\omega)=\chi(\omega,t^*)E(\omega)\, ,
\end{align}
where the susceptibility $\chi(\omega,t^*)$ contains information about the time evolution via $t^*$ of the increasing radius $b(t^*)$.

\begin{figure}[h!!!]
\centering
\includegraphics[scale=0.55, angle=-90]{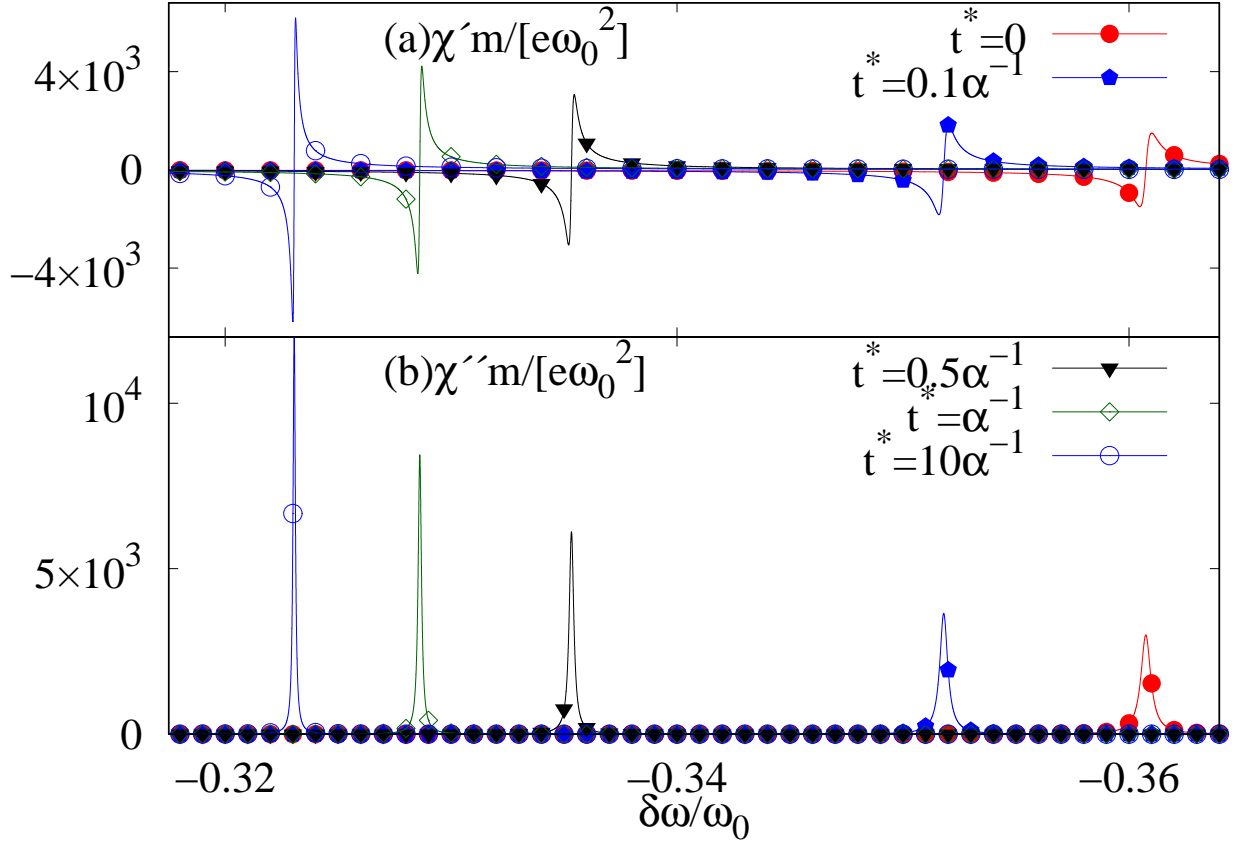}
\caption{Real (a) and imaginary (b) part of the susceptibility $\chi(\omega,t^*)$ to an external field at different times, with $\delta \omega=\omega_m-\omega_0$, $\alpha=10^{-5}\omega_0$,$\tau_{H,D}=10\tau_{B,D}$, $\epsilon_{H,s}=0.1\epsilon_{B,s}$, $\epsilon_{H,\infty}=0.8\epsilon_{B,\infty}$. The sphere extension is $a=3.6${\AA} and $r=0.8a$.\label{FIG3}}
\end{figure}

The real part of the susceptibility is shown in Fig.\ \ref{FIG3}(a) and is connected to the refractive behavior of the molecule, while the imaginary part of the response, shown in Fig.\ \ref{FIG3}(b) is related to its absorptive behavior. 

With increasing time $t^*$ since the onset of the shell formation, the resonances in the absorptive part shifts in frequency to the blue. This results in a smaller relative shift $\delta\omega_m(t^*) = \omega_{m}(t^*)-\omega_0$ of the peak maximum at $\omega_m(t^*)$ with respect to the eigenfrequency $\omega_0$ until the shell is fully formed out. The resonance frequency is smaller than the eigenfrequency $\omega_m(t^*)<\omega_0$ due to the renormalization due to the polar environment.  Thus, the incoming light will be absorbed from the molecule at higher frequencies when the hydration shell around it is more pronounced. This up-shift to the blue of the absorption frequency corresponds to an energy which may be understood as a part of the energy needed to build up the hydrogen-bond network within the hydrophobic hydration shell.

To quantify this work performed, we can use the Gibbs free energy of solvation for our model with $\Delta G=\Delta H -T\Delta S=\Delta E +p \Delta V - T\Delta S$, while $\Delta H$ is the enthalpic change, $\Delta E$ is the change of the internal energy, $p\Delta V$ the volume work and $T\Delta S$ the heat exchange during the solvation. The change of internal energy $\Delta E= \Delta W - p\Delta V+T\Delta S$ contains a contribution of non-expansion work $\Delta W$, while one keeps the pressure and temperature constant. The defined Gibbs free energy then reads $\Delta G=\Delta W$, which reflects the maximal amount of  work that can be extracted from the solute ($\Delta G<0$), or has to be added to the solute ($\Delta G>0$) during solvation in order to letting it dissolve. 
In our case, $\Delta G$ will be positive since we have to perform work to electrostatically orient the water to form a hydration shell. In fact, to be quantitative, we can extract this work from the blue shift of the peaks in the susceptibility, see, e.g., Fig.\ \ref{FIG3}(b). For the parameters used there, we can read off the frequency difference of $\Delta \delta \omega=0.04 \omega_0$, such that $\Delta G=\Delta W=0.04 \hbar \omega_0$.  The performed work stems from the explicitly time-dependent shell thickness performed from outside.  The total required Gibbs free energy $\Delta G>0$ for the whole hydrophobic solvation process is positive, reflecting the low solubility of iodine and thus, the hydrophobic character. The frequency shift is directly measurable in the absorption spectrum of the solute.

In MD simulations of hydrophobic solvation of noble gases the positive free Gibbs energy $\Delta G = \Delta H- T \Delta S>0$ results out of an extremely large loss of entropy and hence a negative $\Delta S$ at room temperature which also have experimental evidence.\cite{sed2011,gar1996} A prominent interpretation is that the water molecules in the hydration shell are more strongly coupled among each other and to the solute which reduces their mobility, and hence, the entropy of the system. Therefore, the entropic penalty for the hydrophobic solvation becomes directly measurable.

\begin{figure}[h!!!]
\centering
\includegraphics[scale=0.6, angle=-90]{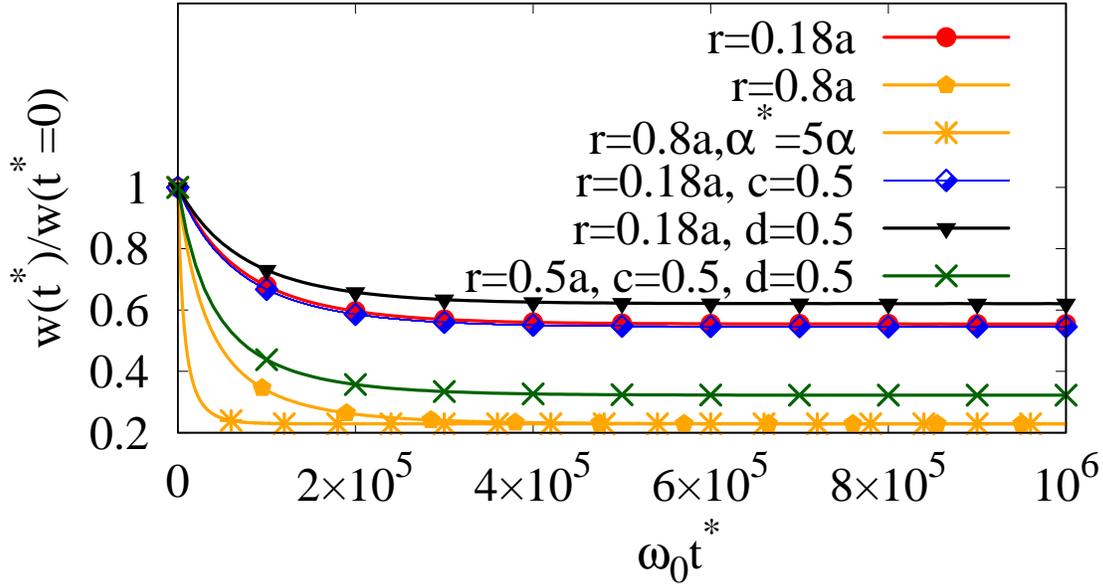}
\caption{Full width at half maximum $w(t^*)/w(t^*=0)$ of the peak in the imaginary part of the susceptibility $\chi(\omega,t^*)$ in dependence of the time $t^*$ elapsed since hydration shell formation has started, for different final shell extensions $r$. The high-frequency dielctric constant is set to $\epsilon_{H,\infty}=c\epsilon_{B,\infty}$ with $c=0.8$, if not stated otherwise. The static dielectric constant $\epsilon_{H,s}=d\epsilon_{B,s}$ with $d=0.1$, if not stated otherwise. The inner Onsager sphere radius reads $a=3.6${\AA} and $\alpha=10^{-5}\omega_0$, $\alpha^*=5\alpha$ and the Debye times are  $\tau_{D,H}=10\tau_{D,B}$.\label{FIG4}}
\end{figure}

Additionally, one observes a strongly reduced linewidth when $t^*$ is growing. This is shown in Fig.\ \ref{FIG3} (b) and, more explicitly, in Fig.\  \ref{FIG4}, and perfectly reflects the reduced damping of the central dipole. The basic physical origin are the less flexible water molecules of the shell and the  strengthened hydrogen bonds of the water molecules, see also Fig.\ \ref{FIG2}. The range of the hydrogen bonds in the shell is smaller as compared to bulk water, which indicates a high structural ordering.  In the Debye model, this effect is included by a smaller static and high-frequency dielectric constant in the shell. \cite{tie2010} The linewidth, measured by the full width at half maximum of the peak in the absorptive part of the response, is more reduced with time $t^*$ if the static dielectric constant is more reduced in the shell, as shown in Fig.\ \ref{FIG4}. The final width is  reached faster, with a more rapid shell formation.

\begin{figure}[h!!!]
\centering
\includegraphics[scale=0.55, angle=-90]{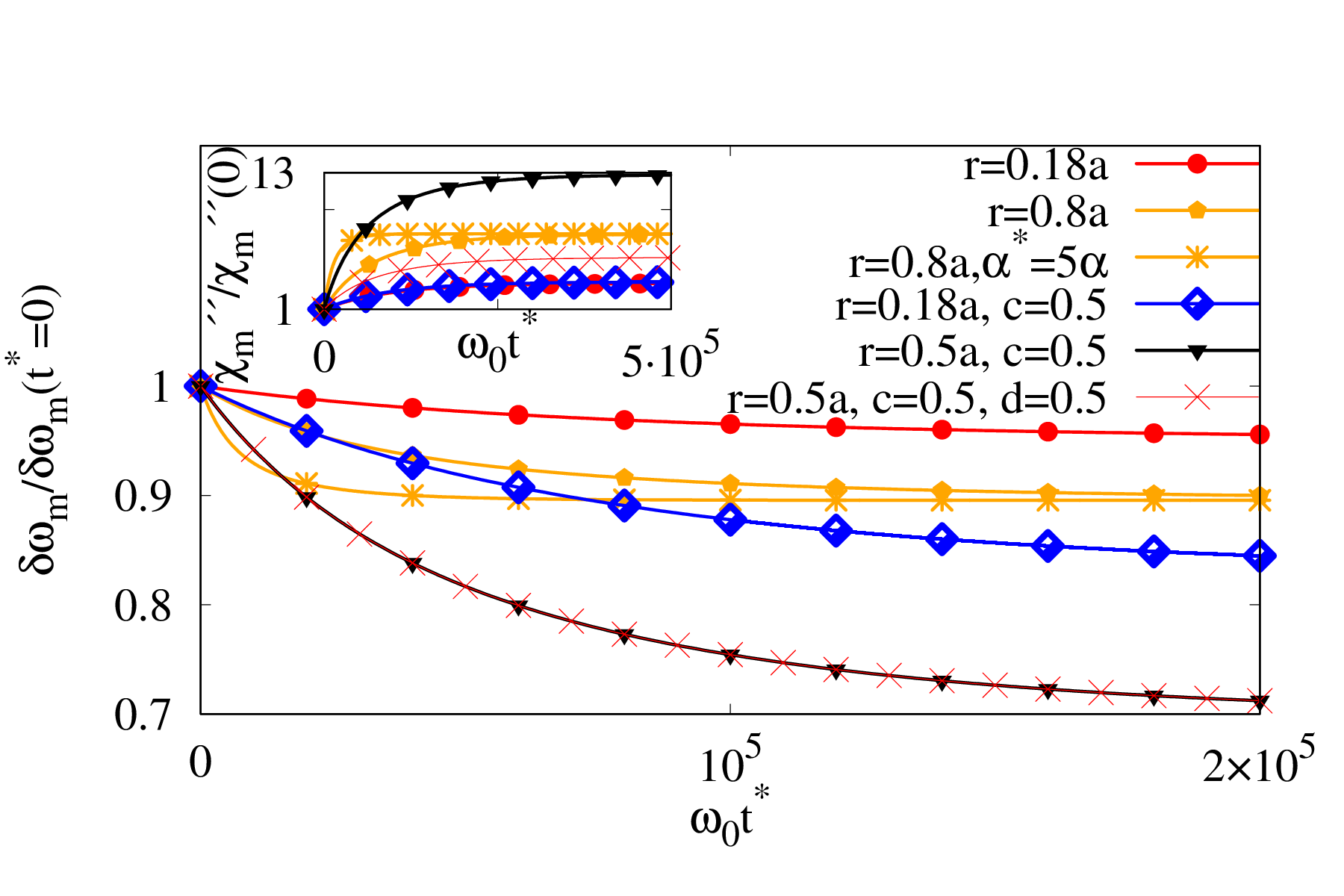}
\caption{Relative frequency shift $\delta \omega_{m}(t^*)=\omega_{m}(t^*)-\omega_0$ of the peak in the imaginary part $\chi_{m}''(t^*)$ in dependence of  $t^*$ for different final shell extensions $r$. The high-frequency dielctric constant is  $\epsilon_{H,\infty}=c\epsilon_{B,\infty}$ with $c=0.8$, if not specified otherwise. The static dielectric constant is $\epsilon_{H,s}=d\epsilon_{B,s}$ with $d=0.1$, if not specified otherwise.  Inset: Evolution of the peak maximum in the imaginary part in dependence of $t^*$. The inner Onsager sphere radius is $a=3.6${\AA},  $\alpha=10^{-5}\omega_0$, $\alpha^*=5\alpha$ and $\tau_{D,H}=10\tau_{D,B}$.\label{FIG5}}
\end{figure}

To be more quantitative, we show in Fig.\ \ref{FIG5} the  total blue shift  $\delta \omega$ of the resonance frequency over time $t^*$. It increases for a larger final shell thickness $r$, as more energy is needed to form the hydrophobic shell. This result coincides with the higher energy required to dissolve hydrophobic molecules with a bigger volume, because a molecule with thicker hydration shell takes up an enlarged volume in the solution \cite{cha2005}. The shift to the blue occurs faster for a quicker shell formation.
Moreover, a strong up-shift is observed with a reduced high-frequency dielectric constant $\epsilon_{H,\infty}$ of the bound water. This  is seen by changing the parameter $c$ and shown in Fig.\ \ref{FIG5}. In contrast, no impact on the frequency shift is found  for a reduced static dielectric constant  $\epsilon_{H,s}$. Analytically, this can be understood via the impact of the high-frequency dielectric constant on the real part of the susceptibility $\xi$ in Eq.\ (\ref{EQ8}) which leads to a frequency renormalization, while the static dielectric constant has an imperceptible impact on $\omega_0$.  

The maximum of the absorption peak increases with time $t^*$ until the hydration shell is finally formed out, which is shown in Fig.\ \ref{FIG5} (inset). This enhanced response is due to the fact that the strongly damping bulk water is more and more replaced by the less damping water more tightly bound in the shell. The final value of the peak maximum is faster reached with a more rapid shell formation. This effect is more pronounced for a thicker shell. The reduced damping and the increase of the absorption maximum  validate the decreased linewidth in Fig.\ \ref{FIG4}. We also find an enhanced growth of the peak maximum  and a smaller line width (shown in Fig.\ \ref{FIG4}) for a smaller static dielectric constant $\epsilon_{H,s}$, while this effect is not visible for a smaller high-frequency dielectric constant, see inset of Fig.\ \ref{FIG5}. We thus can conclude that the high-frequency constant $\epsilon_{H,\infty}$ contributes more to a renormalization of the eigenfrequency $\omega_0$, while the static constant $\epsilon_{H,s}$ strongly influences the damping and hence the absorption maximum of the central dipole moment.
 
\subsection{B. Time-dependent dielectric permittivity}
 
In the second approach, the shell formation is described by a time-dependent decrease of the static $\epsilon_{H,s}$ and the high-frequency $\epsilon_{H,\infty}$ dielectric constants of the dielectric continuum in a layer region  with finite thickness $r$ around the Onsager cavity within a Debye model. Once the ion has been neutralized by a light pulse, a hydrogen network will be gradually formed associated with its specific dielectric constant. During this process, the dielectric constants will progressively decrease from their initial values of bulk water  to the their final values. Within this picture, the small environmental dipole moments arrange themselves successively in a network of hydrogen bonds within the fixed shell thickness $r$. The overall temporal Debye form will be kept unchanged with the fixed relaxation time $\tau_{H,D}=\tau_{B,D}$, see Eq.\ \ref{EQ1}, while the prefactors decrease according to 
\begin{align}
\label{EQ11}
g(t,t_0)=1-h[1-e^{-\alpha (t-t_0)}]\Theta(t-t_0)\, ,
\end{align}
where $1-h$ describes the final magnitude of the dielectric constants within the shell, while $t_0$ is again the time when the shell formation starts. $\alpha$ defines the rate of the decrease of the dielectric constant. The exponential decrease in Eq.\ \ref{EQ11} is chosen because the hydration shell forms out more rapidly at the beginning, before it successively reaches its final dielectric properties.
The model parameter is chosen as $h < 1$ because the more strongly coupled water network of the shell can be described by a reduced dielectric constant and the dipoles in the shell  are less polarizable than bulk water.
 
Again with the same idea as before, we now treat  $g(t)$ adiabatically in Eqs.\ (\ref{EQ3})-(\ref{EQ6}). After Fourier transforming the resulting reaction field, we find 
\begin{align}
\label{EQ12}
R(\omega)&=\frac{\mu(\omega)}{a^3}\frac{2[g(t^*,t_0)-1][2+g(t^*,t_0)\epsilon_B(\omega)]a^3-2[2+g(t^*,t_0)][g(t^*,t_0)\epsilon_B(\omega)-1]b^3}{[g(t^*,t_0)\epsilon_B(\omega)-1][g(t^*,t_0)-1]2a^3-[2+g(t^*,t_0)][2g(t^*,t_0)\epsilon_B(\omega)+1]b^3}\\ \notag
&\equiv \xi(\omega,t_0,t^*)\mu(\omega)=e\xi(\omega,t_0,t^*)q(\omega),
\end{align}
where $\xi(\omega,t_0,t^*)$ contains all information about the adiabatically decreasing dielectric constant within the solvation shell. We have to ensure that  $\alpha\ll \Gamma$, where $\Gamma$ is the damping of the induced reaction field $A(t)$ within the hydration shell. Hence, $g(t)\equiv g(t^*)$ can be treated parametrically. 
As before, the time $t_0=0$ marks beginning of the hydration shell formation, while $t^*$ is the beginning of the evaluation of the response, the lower boundary of the Fourier transform, reflecting the arrival of the external pulse, where $t^*\geq t_0$. 
We are now able to study the response to an external electric field with the reaction field given in Eq.\ (\ref{EQ12}), 
 which we insert into Eq.\ (\ref{EQ10}).
\begin{figure}[h!!!]
\centering
\includegraphics[scale=0.55, angle=-90]{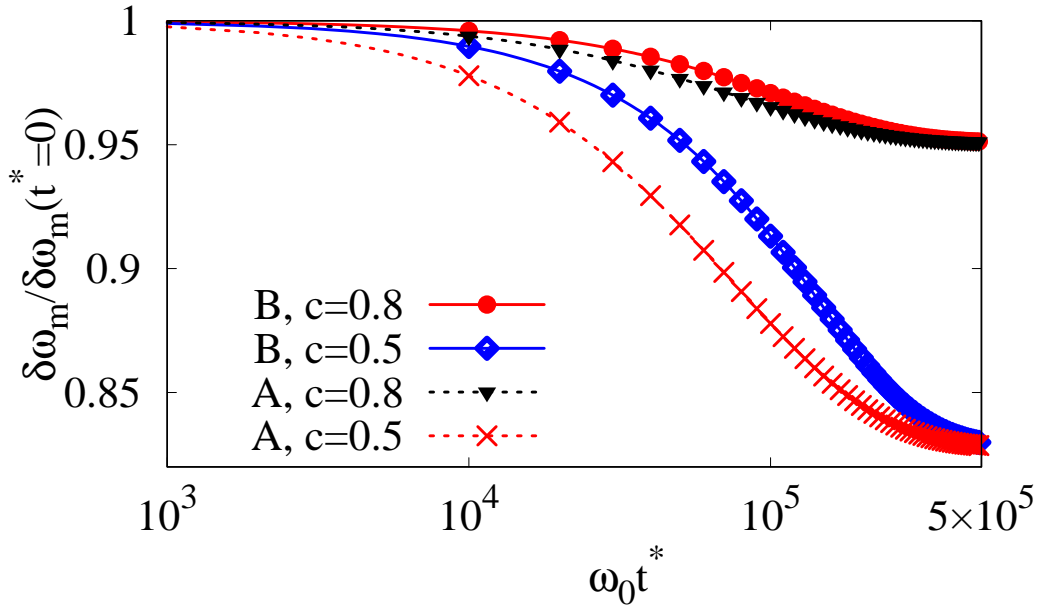}
\caption{Relative frequency shift $\delta \omega_{m}(t^*)=\omega_{m}(t^*)-\omega_0$ of the peak in the imaginary part of the susceptibility $\chi_{m}''(t^*)$ in dependence of the time $t^*$  since hydration shell formation has started, as described by the approach  (A.) with an increasing radius $b(t)$, and by the approach (B.) with a gradual reduction of the dielectric constant in the layer region. The high-frequency and static dielctric constants are $\epsilon_{H,\infty/s}=c\epsilon_{B,\infty/s}$. The inner Onsager sphere radius is $a=3.6${\AA}, the shell thickness $r=0.8a$ and $\alpha=10^{-5}\omega_0$. The Debye times are here $\tau_{D,H}=\tau_{D,B}$ for both scenarios.\label{FIG6}}
\end{figure}
As before with the approach (A.), we also find here a relative frequency up-shift in the absorptive part of the response function. This is shown in Fig.\  \ref{FIG6}, where we compare the results of both approaches. For the dynamically increasing radius, scenario (A.), the frequency shift decreases faster with time than for the decreasing dielectric constant, scenario (B.). Formally, this may be understood by the fact that $b(t)^3$ enters cubic in Eq.\ (\ref{EQ7}) while $g(t)$ appears as  linear in Eq.\ (\ref{EQ12}). 
The surface of the hydration shell grows quadratically in scenario (A.) and thus also the number of involved environmental dipoles, whereas the total number of the involved dipoles in scenario (B.) is constant and they adjust themselves all together at the same time. With a  more reduced dielectric constant in the shell, its polarizability decreases in comparison to the bulk and the total up-shift is more pronounced, as seen in Fig.\ \ref{FIG6}. 

\section{Conclusions}
We have studied the time-resolved response of a molecule, becoming a hydrophobic agent, which is embedded in water as a highly polar environment, around which a layer of hydrated water grows dynamically.  
This model refers to the abstraction of an electron of iodide which results in its neutral but strongly hydrophobic iodine partner  \cite{pha2011}. After the formation of the apolar hydrophobic solute,  hydrogen bonds in the water have to be broken up and the water molecules have to rearrange. This hydrophobic solvation process is accompanied by a formation of a hydration shell which leads to a reduction of the overall entropy such that the solubility is energetically highly unfavorable. In the very proximity of the hydrophobic agent, the polarizability of the water is assumed to be reduced since water molecules form strengthened hydrogen bonds and the fluctuations are slowed down as compared to bulk water. This is modeled by a continuous hydration layer with reduced dielectric constants. 

With this continuum model, we are able to calculate the time-dependent response of the central probe dipole in this nonequilibrium environment to an external electric field. 
Within our calculation, we have assumed that the electric probe pulse is long enough to drive the dipole to a stationary oscillation, but short enough to treat the change of the radius of the cluster adiabatically. In good agreement with the experiment, we find  a  time scale of the shell formation of $\alpha^{-1}\sim \rm ps$ \cite{pha2011,bak2008}. We have studied two scenarios of the hydration shell formation, where, within the first one (A.), the thickness of the spherical shell grows progressively, while, in the second one (B.), the Debye dielectric permittivity  decreases gradually for a fixed shell extension. Both approaches  yield a relative dynamical frequency up-shift of the resonance in the absorptive part of the response function during the shell formation, which signals a positive free energy during the hydrophobic solvation. The frequency blue shift is accompanied by a reduced line width in the absorptive peak which additionally indicates less damping due to a more rigid structural order and stronger hydrogen bonds of the water molecules in the shell as compared to bulk water.  For both scenarios, the dynamic frequency shifts of the absorption resonance reveals the dynamics of the hydration shell formation.
Further experimental investigation by pico- and femtosecond absorption spectroscopy will lead to more detailed insights to elucidate the hydrophobic solvation process on the atomic scale, and, in particular, into the dielectric properties  inside the hydration shell.

\begin{acknowledgement}
This work is supported by the DFG-Sonderforschungsbereich 925 ``Light-induced dynamics and control of correlated quantum systems (project A4). We thank Thomas Elsaesser for illuminative discussions. 


\end{acknowledgement}



\end{document}